\documentstyle[preprint,aps,12pt]{revtex}
\begin{document}
\draft
\hfill\vbox{\baselineskip14pt
            \hbox{\bf ETL-xx-xxx}
            \hbox{ETL Preprint 00-xxx}
            \hbox{June 2000}}
\baselineskip20pt
\vskip 0.2cm 
\begin{center}
{\Large\bf SET based experiments for HTSC materials}
\end{center} 
\vskip 0.2cm 
\begin{center}
\large Sher~Alam$^{1}$,~M.~O.~Rahman$^{2}$,~H.~Oyanagi$^{2}$,
and T.~Yanagisawa$^{1}$
\end{center}
\begin{center}
$^{1}${\it Physical Science Division, ETL, Tsukuba, Ibaraki 305, Japan}\\
$^{2}${\it  GUAS \& Photon Factory, KEK, Tsukuba, Ibaraki 305, Japan}
\end{center}
\vskip 0.2cm 
\begin{center} 
\large Abstract
\end{center}
\begin{center}
\begin{minipage}{14cm}
\baselineskip=18pt
\noindent
  The cuprates seem to exhibit statistics, dimensionality
 and phase transitions in novel ways. The nature of excitations
 [i.e. quasiparticle or collective], spin-charge separation, 
 stripes [static and dynamics], inhomogeneities, psuedogap, effect of 
 impurity dopings [e.g. Zn, Ni] and any other phenomenon in these materials
 must be consistently understood. In this note we suggest Single Electron 
 Tunneling Transistor [SET] based experiments to understand 
 the role of charge dynamics in these systems. Assuming that SET 
 operates as an efficient charge detection system we can expect to 
 understand the underlying physics of charge transport and
 charge fluctuations in these materials for a range of doping. 
 Experiments such as these can be classed in a general
 sense as mesoscopic and nano characterization of cuprates
 and related materials.
\end{minipage}
\end{center}
\vfill
\baselineskip=20pt
\normalsize
\newpage
\setcounter{page}{2}

	In a previous work one of us \cite{alam98} has advanced 
the conjecture that one should attempt to model the phenomena of
antiferromagnetism and superconductivity by using quantum
symmetry group. Following this conjecture to model the phenomenona of
antiferromagnetism and superconductivity by quantum symmetry
groups, three toy models were proposed \cite{alam99-1}, namely,
one based on ${\rm SO_{q}(3)}$ the other two constructed with
the ${\rm SO_{q}(4)}$ and ${\rm SO_{q}(5)}$ quantum groups. 
Possible motivations and rationale for these choices  
were outlined \cite{alam99-1}. In \cite{alam99-2} a model to 
describe quantum liquids in transition from 1d to 2d dimensional 
crossover using quantum groups was outlined. In \cite{alam00-1} 
the classical group ${\rm SO(7)}$ was proposed as a toy model to 
understand the connections between the competing phases and the 
phenomenon of psuedo-gap in High Temperature Superconducting Materials
[HTSC]. Then we proposed in \cite{alam00-2} an idea to
construct a theory based on patching critical points so as to
simulate the behavior of systems such as cuprates.
To illustrate our idea we considered an example 
discussed by Frahm et al., \cite{fra98}. The model
deals with antiferromagnetic spin-1 chain doped with
spin-1/2 carriers. In \cite{alam00-3} the connection between
Quantum Groups and 1-dimensional [1-d] structures such as
stripes was outlined. The main point of \cite{alam00-3}
is to emphasize that {\em 1-d structures play an important
role in determining the physical behaviour [such as the 
phases and types of phases these materials are capable of
exhibiting] of cuprates} and related materials.
In order to validate our quantum group conjecture
for the cuprates, we have considered the connection between 
quantum groups and strings \cite{alam00-4,kak91}.

	The cuprates seem to exhibit statistics, dimensionality
and phase transitions in novel ways. We summarize the following 
interesting features that appear to arise in these materials:
\begin{itemize}
\item{}Pseudogap:-We can consider the reduction
of the density of states near the Fermi energy
as a pseudogap. For example Nakano et al.\cite{nak98}, claim
that magnetic susceptibility measurements of
the cuprate La$_{2-x}$ Sr$_{x}$ Cu O$_{4}$ [LaSCO]
in the T-x phase diagram show two crossover lines
T$_{max}$(x) and T$^{*}$ (x) [where T$_{c}$ $<$ T$^{*}$
$<$ T$_{max}$]. Thus these lines T$_{max}$(x) and T$^{*}$ (x)
are naturally termed the high and low energy pseudogap
respectively. These lines in the T-x diagram
are both montonically decreasing with rising
hole concentration $x$. Below T$_{max}$ magnetic
susceptibility exhibits a broad peak which in the 
usual interpretation is taken to arise from the 
gradual development of the antiferromagnetic spin
correlation. The lower crossover line T$^{*}$
is taken to represent the temperature below which
a spin gap opens up in the magnetic excitation
spectrum around $q=(\pi,\pi)$. 
\item{}Charge-spin separation:-Indications for
electron fractionalization from Angle-Resolved 
Photoemission Spectroscaopy [ARPES] have
been reported \cite{org00}. We note that the Fermi 
liquid is characterized by sharp fermionic
quasiparticle excitations and has a discontinuity in the
electron momentum distribution function. In contrast
the Luttinger liquid is characterized by charge $e$
spin $0$ bosons and spin $1/2$ charge $0$ and the
fermion is a composite of these [i.e. fractionalization].
It is well-known that transport properties
are defined via correlation functions.
The correlation functions of a Luttinger liquid
have a power law decays with exponents that
depend on the interaction parameters. Consequently
the transport properties of a Luttinger liquid
are very different from that of a Fermi liquid.
Photoemission experiments on Mott insulating oxides
seems to indicate the spinon and holon excitations
of a charge Luttinger liquid. However the experimental
signatures of Luttinger liquid are not totally
convincing. To this end we propose SET based experiments
to determine the Luttinger liquid behaviour of
the cuprates, see below.  
\item{}d-wave symmetery:-Experimental evidence
for predominantly d-wave pairing symmetry in both hole- 
and electron-doped high T$_{c}$ cuprate superconductors
has been reported by C.C. Tsuei and J.R. Kirtley \cite{tsu00}.
\item{}Stripes:-For recent overview see abstracts
of Stripes 2000 conference.
\item{}1/8 problem:-The recent experimental
work of Koike et al.~\cite{koi00} indicates that the
dynamical stripe correlations of holes and spins exist
in Bi-2212, Y-123 and also La-124 and that they tend
to be pinned by a small amount of Zn at 
$p \sim 1/8$\footnote{where p is the hole concentration per Cu}, 
leading to 1/8 anomaly.
\end{itemize} 

	Keeping the properties of the cuprates and related 
materials in mind, in this short note we turn our attention to the
possibility of  Single Electron Tunneling Transistor [SET] based 
experiments which can probe the charge dynamics in HTSC cuprates. 
In particular the detection of charge-rich and 
charge-poor [i.e. stripes] and the important question of detection 
of fractional charge $q_{f}~e$ carried by Luttinger excitation
\footnote{The Luttinger excitation is fractionalized
and the elementary excitation carries the fractional
charge $q_{f}~e$ instead of the quantum of charge
$e$ of the electron}.

\begin{itemize}
\item{}SET coupled to HTSC Josephson junction:-The
SET works on Coulomb energy and on the process of 
tunneling. We can simply define a SET transistor as 
a small conductor, usually a small metallic island, 
placed between two bulk external electrodes, that 
forms two tunnel junctions with these electrodes 
\cite{ave00}. The Coulomb charging
of the island takes place due to electron tunneling.
The current $I$ in this system depends on the
electrostatic potential of the island which in turn
is controlled by external gate voltage $V_{g}$.
The SET transistor operation as a detector entails
the measurement of the variations of the voltage  
$V_{g}$ which is sensitive to the current $I$.
It is natural to consider the SET as a quantum 
detector \cite{ave00}. The SET transistor is the
natural measuring device for the potential quantum
logic circuits based on the charge states of 
mesoscopic Josephson junctions \cite{ave00}.
It is natural to consider the SET as a quantum 
detector if one carefully consider its noise 
properties \cite{ave00}. We propose to couple
the SET transistor to HTSC Josephson junction
and study the charge dynamics of the HTSC
materials for various levels of doping. 
The `insulating' layer in the Josephson 
junction is chosen for a particular value of
doping. For example, in LASCO system it is known
that that superconductivity is suppressed
to some extent at hole concentration per Cu
of 1/8. This is considered to be due to
existence of stripes. One would expect measurable 
change in SET current as one goes from this region
towards the purely AF-phase and purely superconducting. 
In particular it would be interesting to explore the 
transition regions of the T-x phase diagram.

\item{}SET transitor coupled to `HTSC material' 
SET transistor:- In this set-up we propose to 
couple a SET transitor to a `HTSC material' 
SET transistor. In the simplest case the
`metallic' island in ordinary SET is replaced
with `HTSC material' for a value of doping
which is in the region of T-x phase diagram
which corresponds to metallic phase and
`strange' metallic [SM] phase. The
SM phase is ascribed to coexistence of
superconductivity and stripe phases. In this region
the material, if it were perfectly oriented
would be a superconductor in one direction
and a strongly-correlated insulator in
the other. Thus one can quantify the SM and
metallic phases of HTSC materials in
detail by using charge transport properties
measured with SET.  

\item{}Single-Cooper pair box:- Tsai et al. \cite{tsa00}, 
have recently demonstrated the time and energy domain
response of an artificially constructed two-level 
system, which is expected to form one of the
possibilities for the basic bit of quantum computing
[Qubit]. This device which has submicron size allows
one to observe quantum coherent oscillations in a
solid state system whose quantum states involved
a macroscopic number of quantum particles. 
As already mentioned it has already  been noted
\cite{ave00} that the SET is a natural measuring device 
for the potential quantum logic circuits based on the 
charge states of mesoscopic Josephson junctions, 
such as Single-Cooper pair box of Tsai et al.~\cite{tsa00}. 
Keeping in mind that HTSC materials are {\em doped} Mott insulators
[unlike the low-temperature superconductors], and
consequently their superconductivity depends on
the level of doping as is clear from the T-x phase
diagram, so that one has underdoped, optimally
doped and overdoped regions. It would be interesting
to think of an experiment that can give us a detailed
look at the effect of changing T$_{c}$ [as doping
is varied] on the charge transport. A possible
experimental set-up could consist of SET transistor
coupled to a HTSC single-Cooper pair box.
\end{itemize}

	To characterize the transport properties in striped
phase in above materials, the experiments must be calibrated
against some standard, which clearly shows the one-dimensional
transport behaviour and also must be closely related to
the cuprate superconductors. One such material is
La$_{1.4-x}$ Nd$_{0.6}$Sr$_{x}$CuO$_{4}$ (x=0.1,0.12,0.15)
which is stripe-ordered non-superconducting relative
of HTSC cuprates.

	In conclusion we have suggested SET based experiments
to characterize	and understand the underlying physics of charge 
transport and charge fluctuations in these materials for a range 
of doping. Experiments such as these can be classed in a general
sense as mesoscopic and nano characterization of cuprates
and related materials.
 
\section*{Acknowledgments}
The Sher Alam's work is supported by the Japan Society for
for Technology [JST].

\end{document}